\documentclass{article}

\usepackage{arxiv}

\usepackage[utf8]{inputenc} 
\usepackage[T1]{fontenc}    
\usepackage{hyperref}       
\usepackage{url}            
\usepackage{booktabs}       
\usepackage{amsfonts}       
\usepackage{nicefrac}       
\usepackage{microtype}      
\usepackage{lipsum}
\usepackage{graphicx}
\graphicspath{ {./images/} }
\usepackage{booktabs}
\usepackage{multirow}
\usepackage{array}
\newcolumntype{C}[1]{>{\centering\arraybackslash}p{#1}}
\usepackage{authblk}
\usepackage{etoolbox} 
\usepackage{xcolor}


\title{Transforming Role Classification in Scientific Teams Using LLMs and Advanced Predictive Analytics}

\author[1]{Wonduk Seo} 
\author[1,2*]{Yi Bu}

\affil[1]{\emph{Department of Information Management, Peking University,    Beijing 100871, China}}
\affil[2]{\vspace{1.5em}\emph{Peking University Chongqing Research Institute of Big Data, Chongqing 401332, China}}
\affil[*]{\textbf{Corresponding author: Yi Bu} $($\texttt{buyi@pku.edu.cn}$)$}

\begin{document}
\maketitle

\begin{abstract}
Scientific team dynamics are critical in determining the nature and impact of research outputs. However, existing methods for classifying author roles based on self-reports and clustering lack comprehensive contextual analysis of contributions. Thus, we present a transformative approach to classifying author roles in scientific teams using advanced large language models (LLMs), which offers a more refined analysis compared to traditional clustering methods. Specifically, we seek to complement and enhance these traditional methods by utilizing open source and proprietary LLMs, such as GPT-4, Llama3 70B, Llama2 70B, and Mistral 7x8B, for role classification. Utilizing few-shot prompting, we categorize author roles and demonstrate that GPT-4 outperforms other models across multiple categories, surpassing traditional approaches such as XGBoost and BERT. Our methodology also includes building a predictive deep learning model using 10 features. By training this model on a dataset derived from the OpenAlex database, which provides detailed metadata on academic publications—such as author-publication history, author affiliation, research topics, and citation counts—we achieve an F1 score of 0.76, demonstrating robust classification of author roles.
\end{abstract}


\textbf{Wonduk Seo}: \texttt{0009-0008-6070-1833};
\textbf{Yi Bu}: \texttt{0000-0003-2549-4580}

\section{Introduction}
In the dynamic evolution of scientific research, the structure of research teams plays a significant role in shaping the nature and impact of their outputs~\cite{wuchty2007increasing}. In such settings, accurately classifying the roles of team members is critical to understanding the mechanisms that drive scientific innovation and productivity. Effective role classification not only helps to recognize individual contributions, but also improves the management and optimization of research teams. Traditionally, methods have focused on categorizing authors into leadership, direct support, and indirect support. Leadership roles typically involve tasks such as designing and directing research, direct support encompasses activities like data collection and analysis, and indirect support includes providing feedback and editing manuscripts. The motivation behind focusing on these three categories is to understand the hierarchical structure and dynamics within scientific teams. Leadership roles significantly influence the direction and impact of a project, and identifying these roles allows to analyze patterns such as how leadership distribution affects team performance and the overall success of collaborative research efforts. Those are categorized based on clustering using terms derived from self-reported data, which refers to the information provided by the authors themselves about their contributions to a paper~\cite{xu2022flat}.

While this approach has yielded invaluable insights, including metrics such as the L-ratio, the proportion of leadership roles within teams, they are limited in their ability to capture the full context and nuances of authors' contributions. Existing methods often lack the depth required to understand the specific context and impact of individual contributions. Furthermore, clustering used in previous research can be limited by their reliance on static data, failing to adapt to the evolving nature of scientific collaboration. For instance, an author who contributed to research design, data analysis, and manuscript writing might be grouped under one general category without distinguishing the importance of each task. This results in a static and incomplete representation of an author's role and impact on the research project. In addition, term-based clustering does not account for the complexity and multifaceted nature of authors' contributions, as it does not differentiate between the importance or context of multiple contributions made simultaneously by an author. This oversimplification can obscure the depth of individual effort and intellectual contribution, leading to a limited and sometimes misleading understanding of team dynamics and research contributions. One of the main challenges in classifying author roles lies in the complexity and nuance of contribution statements. Authors often describe their contributions using sophisticated and context-specific language, making it difficult for traditional models to accurately interpret and categorize these roles. This complexity requires a deep understanding of context and semantics that goes beyond simple keyword matching or basic natural language processing.

Given the significant advances in large language models (LLMs), they have been widely adopted for various science of science tasks such as identifying research trends, categorizing scholarly articles, extracting key information, and mapping the evolution of scientific fields~\cite{xu2024ai,krenn2020predicting}. Our research explores a different setting: the use of LLMs such as GPT-4, Llama3 70B, Llama2 70B, and Mistral 7x8B for role classification in scientific research~\cite{achiam2023gpt,jiang2024mixtral,touvron2023llama}. These advanced models can analyze patterns from the context of papers and authors, providing a more complete and dynamic understanding of team structures and collaboration dynamics, compared to traditional methods that often fail to capture the depth and context of individual contributions because they rely on static data and predefined categories. Furthermore, LLMs can perform few-shot learning, effectively adapting to new tasks with minimal examples, which is particularly advantageous when labeled data is scarce. They can handle the disambiguation of roles by understanding the interplay of different contributions within the same context, differentiating between leadership and support roles even when an author is involved in multiple tasks. By leveraging LLMs, we aim to overcome this limitation and provide deeper insights into the intricate roles within scientific teams.

We present a comprehensive framework that utilizes such LLMs to classify roles within scientific teams. Our methodology includes a few-shot prompt to accurately categorize author roles. Furthermore, after the LLM inference step, we aim at feature engineering: the framework incorporates ten extracted features reflecting each author's contributions and characteristics, such as contribution to references, contribution to topics, probability of leading, probability of managing correspondence, career age, citation count, unique topics, total publications, citation impact per year, and institutional diversity, by using a dataset derived from the OpenAlex database, which provides extensive metadata on academic publications, including citation counts, author affiliations, research topics, and publication history~\cite{priem2022openalex}. We further train a deep learning model using 10 extracted features to classify author roles with high accuracy. Moreover, we conducted SHAP (SHapley Additive exPlanations)~\cite{lundberg2017unified} analysis to investigate the importance of these features in the prediction tasks, providing valuable insights into their contributions. This approach not only complements, but also improves upon traditional methods, providing a more detailed and accurate classification of author roles. By leveraging the advanced capabilities of LLMs, our research aims to transform the landscape of scientific team role classification and provide a more nuanced understanding of team structures and collaboration dynamics.

\section{Related Works}
\subsection{Dynamics of Scientific Teams}
The structure and dynamics of scientific teams have been studied extensively, revealing a growing dominance and importance of team science in knowledge production. Research shows that teams tend to produce more highly cited research than individuals, a trend that has increased over time~\cite{wuchty2007increasing}. Teams now produce more high-impact research, a distinction once dominated by individual authors~\cite{uzzi2013atypical}. Katz and Martin~\cite{katz1997research} emphasized that collaboration among scientists leads to greater pooling of resources, sharing of knowledge, and combining of expertise, which in turn improves the quality and efficiency of scientific research. The interdisciplinary nature of many scientific teams allows for innovative approaches and solutions to complex problems~\cite{melin1996studying}.

Xu et al.~\cite{xu2022flat} introduced the concept of the L-Ratio (Leadership Ratio), which measures the proportion of leadership roles within teams in their study of team structures, and demonstrated that flat team structures, which encourage equal contributions and a less hierarchical organization, are critical for fostering scientific innovation and improving team performance and research outcomes. Their findings are consistent with previous research suggesting that less hierarchical teams foster a more open exchange of ideas and promote creative problem solving~\cite{hollenbeck2012beyond,woolley2010evidence}. Furthermore, Wu et al.~\cite{wu2019large} found that smaller teams tend to produce more disruptive research compared to larger teams, which generally advance existing scientific knowledge. This indicates the critical role of team size in influencing research outcomes.

In addition, Cummings and Kiesler~\cite{cummings2005collaborative} suggested that team diversity, both in terms of disciplinary background and geographic location, is a significant predictor of team innovation and productivity. This diversity brings a variety of perspectives and skills to the table, leading to richer and more innovative outcomes. Moreover, Anicich et al.~\cite{anicich2015hierarchical} demonstrated that hierarchical cultural values predict success and mortality in high-stakes teams. However, they also noted the challenges associated with coordinating diverse teams, highlighting the need for effective communication and management strategies to realize the full potential of team diversity. Haeussler and Sauermann provided insights into how the division of labor within teams is influenced by factors such as team size and interdisciplinarity, emphasizing its impact on collaborative knowledge production~\cite{haeussler2020division}.

\subsection{Traditional Methods of Role Classification}
Traditionally, in the field of information science, clustering methods have been used to classify academic articles into research topics based on citation relationships and keyword analysis. Waltman and Eck~\cite{waltman2012new} developed a methodology for constructing a publication-level classification system that uses citation networks and keyword co-occurrence to identify and group related research topics. This approach allows for a more objective and data-driven classification of research outputs, providing insights into the structure and evolution of scientific fields. Similar clustering techniques have been used in several other domains to categorize entities based on relational data. For example, bibliometric analysis and network clustering have been used to identify influential authors and research trends within specific disciplines~\cite{boyack2010co}. These methods often employ algorithms such as community detection and modularity optimization to uncover hidden patterns in complex datasets~\cite{blondel2008fast}. By using these techniques, researchers can gain a deeper understanding of the collaborative networks and intellectual landscapes that underpin scientific progress. Additionally, Glänzel and Schubert~\cite{glanzel2004analysing} discussed the use of bibliometric methods to map the structure and dynamics of research fields.

However, an increasing focus has emerged on classifying roles within academic research teams, recognizing the importance of understanding team dynamics and individual contributions. Recent studies have proposed frameworks for identifying different types of research teams and their roles. For instance, a study introduced algorithms for team identification that categorize members into project-based, individual-based, and representative groups based on their contributions and collaboration patterns~\cite{cheng2024method}. Another work emphasized the need for a unified system to classify essential team roles beyond traditional task and social categories~\cite{fujimoto2016team}. In addition to these quantitative techniques, qualitative approaches have also been used to classify roles within academic research. Conger~\cite{conger1998qualitative} argued that qualitative research is essential for understanding complex phenomena such as leadership within research teams. These qualitative methods complement quantitative approaches, providing a more complete picture of collaborative processes and individual contributions. Holbrook~\cite{holbrook2017peer} highlighted the importance of qualitative evaluations in understanding the context and impact of research contributions. Moreover, different traditions of author ranking significantly shape the perception and classification of contributions. In many fields, the first author is typically seen as the primary contributor, while in others the last author, often the senior researcher or principal investigator, carries more prestige~\cite{endersby1996collaborative}. These different conventions have implications for the assignment and recognition of roles. Additionally, systems such as the Contributor Roles Taxonomy have been developed to provide a transparent and detailed description of individual contributions~\cite{brand2015beyond}. It specifies roles such as conceptualization, methodology, and software development, allowing for accurate attribution of credit.

Recently, metrics such as the L-Ratio have been used to measure the weight of leadership roles within research teams, providing a quantitative approach to assess the influence and contributions of different team members~\cite{xu2022flat}. This metric has shown that flat and egalitarian teams tend to produce more novel ideas compared to tall, hierarchical teams, highlighting the importance of team structure in fostering innovation. These findings emphasize the role of team dynamics and structure in driving creative and innovative research outcomes.

\subsection{Large Language Models (LLMs) and Application in Science of Science}
The advent of large language models (LLMs) such as GPT-4, Llama, and Mistral models has brought significant advances in text analysis and the production of human-like text, making them valuable tools for role classification in scientific research~\cite{brown2020language}. Recent studies have demonstrated the effectiveness of LLMs in several domains, including scientific text analysis, classification tasks, and information extraction~\cite{xu2024ai}. The development of these models can be traced back to the introduction of transformers, a groundbreaking architecture introduced in the paper \emph{Attention is All You Need}~\cite{vaswani2017attention}. Transformers use self-attention mechanisms to process input data in parallel, allowing models to capture long-range dependencies and contextual relationships more effectively than traditional recurrent neural networks (RNN)~\cite{sherstinsky2020fundamentals}. In addition to the transformer architecture, several encoder-decoder models have emerged that exploit the strengths of transformers for different applications. In particular, BERT (Bidirectional Encoder Representations from Transformers), RoBERTa~\cite{liu2019roberta}, and T5 (Text-to-Text Transformer) have been influential. BERT focuses on pre-training a deep bidirectional transformer by predicting masked words in a sentence, allowing it to capture complex context~\cite{kenton2019bert}, while T5 frames all natural language processing (NLP) tasks as text-to-text problems, enabling the model to learn a variety of tasks through a consistent training framework~\cite{raffel2020exploring}. Decoder-only models, such as the GPT series, have also had a profound impact on the field, using a transformer decoder architecture to generate text by predicting the next word in a sequence, given all previous words~\cite{radford2018improving}.

While BERT and its variations have been widely applied in the science of science, the use of LLMs in this field is still emerging. Recent studies have begun to explore the potential of LLMs in evaluating research quality and classifying research aims. For instance, Thelwall et al. investigated the effectiveness of ChatGPT in assessing research quality under various conditions and examined the capability of ChatGPT to identify journal article quality across different academic disciplines~\cite{thelwall2024evaluating,thelwall2024fields,wu2024scientific}. In the context of scientometrics, LLMs are used for various tasks such as identifying research trends, categorizing scholarly articles, extracting key information, and mapping the evolution of scientific fields. These capabilities enable researchers to automate the analysis of large volumes of scientific literature, facilitating more efficient and comprehensive studies of research impact and scholarly networks~\cite{xu2024ai}. For instance, LLMs can assist in generating literature reviews by summarizing key points from numerous papers, thus saving researchers significant time and effort~\cite{aydin2022openai}. Additionally, they can help identify emerging research topics and influential authors by analyzing citation patterns and keyword co-occurrences~\cite{krenn2020predicting}.

Another important application of LLMs in scientometrics is their role in improving the transparency and reproducibility of research. By providing automated tools for literature review and data analysis, LLMs can help ensure that research methods and results are consistently documented and easily accessible. This can facilitate peer review and replication studies, which are critical to validating scientific findings~\cite{gilardi2023chatgpt}. In addition, the use of LLMs to automate the synthesis of research findings can lead to more comprehensive meta-analyses, which are essential for drawing generalized conclusions from multiple studies~\cite{wang2023can}. These advances contribute to a more robust and reliable scientific knowledge base. Furthermore, LLMs are increasingly being used to democratize access to scientific knowledge. By providing sophisticated text analysis and summarization tools, LLMs enable researchers from under-resourced institutions or regions to more effectively engage with the latest scientific literature. This can help bridge the gap between well-funded research institutions and those with fewer resources, promoting a more equitable distribution of scientific knowledge and opportunities for collaboration~\cite{rivas2023marketing}.

\section{Dataset}
We first utilize self-reported data collected from papers published in prominent journals such as PNAS, Nature, Science, and PLoS One from 2003 to 2020, which was opened by Xu et al~\cite{xu2022flat}. These four journals were selected due to their high impact and influence in the scientific community, ensuring high quality and significant research data. In addition, recent efforts to make scientific research fairer and more transparent have led these leading journals to require detailed reporting of each author's specific contributions. This requirement helps ensure that everyone involved in the research receives proper credit, making these journals ideal for our analysis. Specifically, PLoS One was included as it mandates comprehensive author contribution statements similar to the other journals, providing rich and structured data for analysis. Additionally, its broad scope across various scientific disciplines enables a more diverse representation of research team structures. To ensure consistency and reduce selection bias, we followed the methodology of Xu et al.~\cite{xu2022flat} and stratified our sampling across these four journals. This selection provides a representative dataset that supports future work to understand and improve the dynamics of scientific collaboration.

The dataset focuses primarily on the "Author Contribution" field, which contains detailed information about each author's role and contributions to the paper. This field provides insight into various aspects of authors' participation in the research process, such as their involvement in experimental design, data analysis, manuscript writing, and other key tasks. By analyzing this detailed self-reported data, we were able to accurately classify authors' roles into the categories of Leadership, Direct Support, and Indirect Support. For instance, Leadership roles involve tasks such as designing and directing the research, Direct Support includes helping with data collection and analysis, and Indirect Support includes activities such as providing feedback and editing the manuscript.

In addition, for feature engineering purposes, we use the OpenAlex database, which is a rich resource that provides extensive metadata on academic publications, including citation counts, author affiliations, research topics, and publication history~\cite{priem2022openalex}. This database allowed us to expand our dataset by incorporating more comprehensive academic profiles and publication records into our analysis, while ensuring that the data was clean and consistent. This included standardizing author names, resolving ambiguities in author identities, and ensuring that all relevant metadata from OpenAlex were correctly linked to the self-reported data.

In more detail, our dataset consists of two main parts for evaluation and modeling. In the first step, for the LLM evaluation, we used a dataset containing papers from 5,000 distinct authors, covering papers published between 2003 and 2020. To ensure a representative sample and maintain quality for validation purposes, we selected 250 papers from each of four prominent journals—PNAS, Nature, Science, and PLoS One. This selection process focused on papers with team sizes ranging from 2 to 8 authors, as these typically offer a clear division of labor and contribution roles. Through data cleaning and matching processes, we successfully aligned approximately 97\% of our selected papers to their corresponding records in OpenAlex. The small proportion of papers that could not be mapped were due to inconsistencies in metadata, such as variations in author names or missing identifiers. In the second step, for feature engineering and modeling, we expanded our dataset to 2,000 papers from the same time period by using GPT-4, with 500 papers from each of the four journals. The average team size in both parts is consistently five authors per paper.

\section{LLM-Based Role Classification: Methods and Evaluation}
In this stage, we focus on classifying and categorizing roles within scientific teams using self-reported data. Our methodology involves the use of open-source instructional LLMs and the GPT-4 model, augmented with few-shot prompting. Through this process, we aim to achieve a detailed categorization of scientific roles, thereby improving our understanding of team dynamics in scientific research environments. This  classification serves as a crucial foundation for subsequent predictive modeling. While LLMs provide high-quality classifications, their computational demands make them impractical for large-scale applications. To address this, the LLM-generated labels will be leveraged to train a predictive model capable of efficiently scaling to extensive datasets. Section~\ref{sec:role-classification-predictive-model} details this predictive modeling process, which extends our role classification framework to broader scientific collaborations.

\subsection{Overview of LLM-Based Role Classification Tasks}
\begin{figure}[h]
    \centering
    \includegraphics[width=0.8\textwidth]{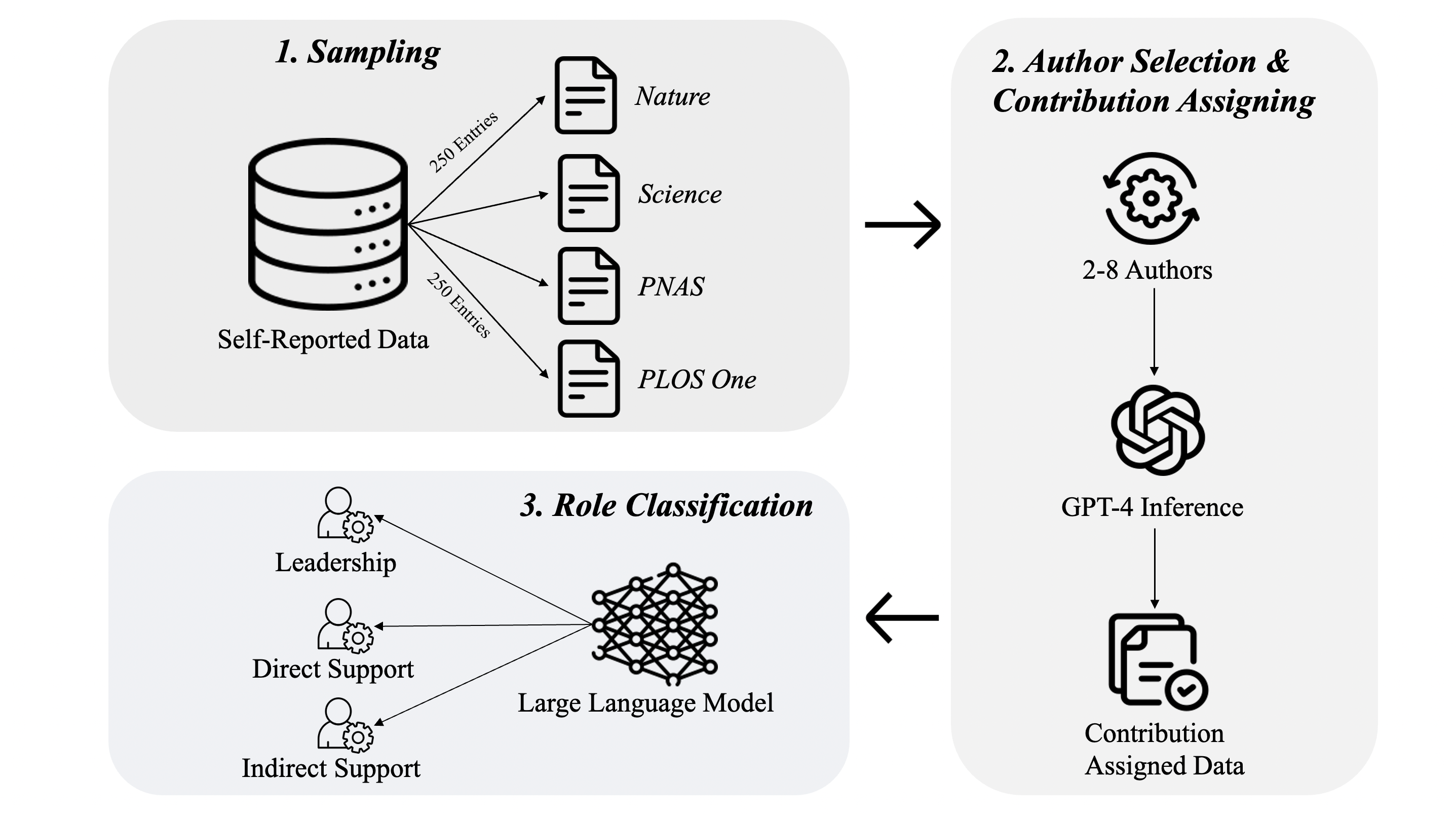} 
    \caption{Illustration of the workflow involving data sampling, preprocessing, contribution assigning, and classification task for the LLM-based role classification.}
    \label{fig:workflow1}
\end{figure}

As shown in Figure~\ref{fig:workflow1}, to ensure a representative sample and maintain quality for validation purposes, we selected 250 entries from each journal. This selection process focused on papers with between 2 and 8 authors, as these typically offer a clear division of labor and contribution roles.

Using GPT-4, we assigned specific contributions to each author, categorizing them as Leadership, Direct Support, or Indirect Support based on predefined criteria. This step involved analyzing the "Author Contribution" field of each paper, which details each author's participation in various research activities such as experimental design, data analysis, manuscript writing, and other key tasks. Using the advanced natural language understanding capabilities of GPT-4, we accurately assigned contributions and categorized roles.

At the end of this assignment process, we compiled a dataset of 5,000 rows. We ensured this by first constraining the number of authors in each paper to be between 2 and 8 and randomly sampling 250 papers from each of the 4 journals, resulting in a total of 1,000 selected papers. Given that the average number of authors per paper was 5, we expanded the dataset to achieve a total of 5,000 rows:
\[
R = J \times E \times \bar{A}
\]
where:
\begin{itemize}
    \item $J = 4$ (number of journals);
    \item $E = 250$ (number of entries selected from each journal); 
    \item $\bar{A} = 5$ (average number of authors per paper).
\end{itemize}
GPT-4 proved to be effective in accurately assigning papers and contributions, ensuring a balanced and comprehensive dataset for analysis.

\subsection{Prompt Engineering for Role Classification}
Once the contributions were assigned, we evaluated the intrinsic capabilities of the non-fine-tuned LLMs through few-shot prompt engineering. Our focus was on evaluating the performance of these models in accurately classifying roles based on predefined criteria. Each role category -- Leadership, Direct Support, and Indirect Support -- was evaluated and our expectations for each classification were outlined.

\begin{figure}[h]
    \centering
    \includegraphics[width=0.8\textwidth]{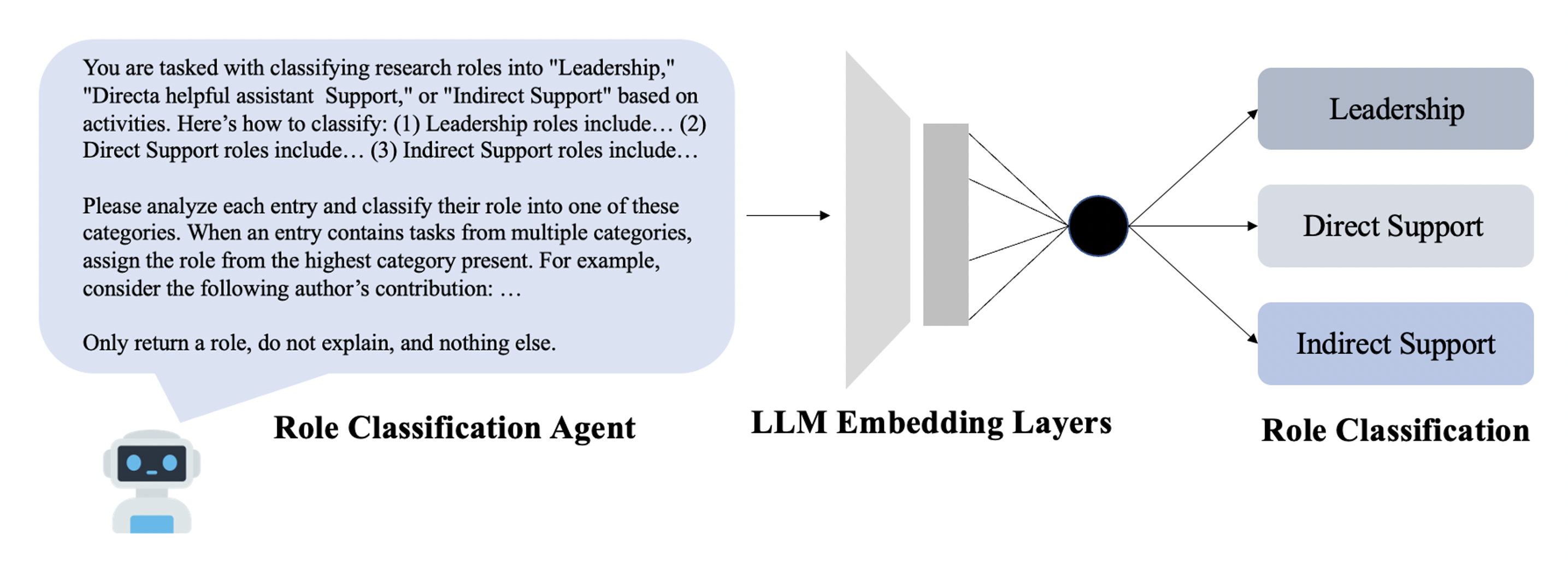} 
    \caption{Illustration of the workflow involving prompt generation, LLM inference, and classification results.}
    \label{fig:workflow2}
\end{figure}

To facilitate this, we created specific prompts for the models, guiding them to categorize and classify research roles based on the activities associated with each role, as illustrated in Figure~\ref{fig:workflow2}. We also explored zero-shot prompting, where the model relies solely on its pre-trained knowledge without additional examples. However, we observed that the model's predictions in zero-shot mode were less stable, often producing inconsistent or ambiguous role assignments. Without explicit demonstrations, LLMs sometimes misclassified contributions, particularly when role descriptions were subtle or overlapping. In contrast, our few-shot approach, which provided targeted examples, significantly improved classification accuracy by reducing ambiguity and ensuring better role differentiation. 

To construct effective prompts, we defined a structured set of role categories and associated key responsibilities, ensuring clarity in classification. These role definitions provided the necessary context for LLMs to accurately assign contributions, as detailed below:

\begin{enumerate}
    \item \textbf{Leadership}: Responsibilities include designing, conceptualizing, directing, supervising, coordinating, interpreting, conducting, and writing the research.
    \item \textbf{Direct Support}: Tasks include helping, assisting, preparing, collecting, and analyzing.
    \item \textbf{Indirect Support}: Tasks include participating, providing, contributing, commenting, editing, and discussing.
\end{enumerate}

Given a dataset detailing the tasks of authors in a research project, the models were instructed to analyze each entry and assign roles accordingly. If an entry contained multiple categories, the role was selected from the highest category present. For example, if an author's activities included both "designing" (a Leadership task) and "providing" (an Indirect Support task), the author was classified under "Leadership" because it ranks higher in the role hierarchy. This structured hierarchy ensured that only one role was assigned to each author.

\subsection{LLM Model Configuration and Performance Comparison}
Initially, we manually labeled corresponding roles for each author, which provided a benchmark for subsequent automated classification by following the criteria outlined in our prompts, assigning roles based on specific activities. We carefully examined each contribution statement, considering multiple facets such as the nature of the tasks performed, the level of responsibility, and the overall impact on the research. To automate this process, we used LLMs for inference, selecting four models: GPT-4, Llama3 70B, Llama2 70B, and Mistral 7x8B. For each model, we standardized the prompting parameters to ensure consistency across evaluations. Specifically, we set temperature to 0.01 to minimize randomness in responses and provide more deterministic results. This setup allowed us to directly compare each model in accurately replicating the human-assigned roles. We performed a detailed performance comparison between GPT-4, Llama3 70B, Llama2 70B, and Mistral 7x8B on three key metrics: F1 score, precision, and recall.

Our results demonstrated the superior ability of GPT-4 to accurately classify roles. As shown in Table~\ref{tab:table1}, GPT-4 consistently led in precision and recall across all role categories, making it the most reliable of the models evaluated. Specifically, in the Leadership classification, GPT-4 achieved a precision of 0.995 and a recall of 0.991, significantly outperforming Llama3 70B, which had a precision of 0.996 but a lower recall of 0.851. For Llama3 70B, while excelling in precision for leadership, it did not perform as well in recall compared to GPT-4. Llama2 70B and Mistral 7x8B have shown moderate to poor performance in all metrics.

\begin{table}[h]
    \centering
    \caption{Comparison of Model Performance on Role Classification Tasks.}
    \label{tab:table1}
    \begin{tabular}{
        >{\centering\arraybackslash}p{1.5cm}  
        >{\centering\arraybackslash}p{3cm}    
        l                                       
        c                                       
        c                                       
        c                                       
    }
        \toprule
        \textbf{Index} & \textbf{Model Name} & \textbf{Role} & \textbf{F1-Score} & \textbf{Precision} & \textbf{Recall} \\
        \midrule
        \multirow{4}{*}{1} & \multirow{4}{*}{GPT-4-1106} 
            & Leadership         & 0.993 & \textbf{0.995} & 0.991 \\
         &  & Direct Support      & 0.950 & 0.945 & \textbf{0.953} \\
         &  & Indirect Support    & 0.947 & 0.925 & \textbf{0.970} \\
         &  & Macro Avg.          & \textbf{0.963} & 0.955 & 0.972 \\
        \midrule
        \multirow{4}{*}{2} & \multirow{4}{*}{Llama3 70B} 
            & Leadership         & 0.918 & \textbf{0.996} & 0.851 \\
         &  & Direct Support      & 0.569 & 0.414 & \textbf{0.911} \\
         &  & Indirect Support    & 0.841 & 0.748 & \textbf{0.961} \\
         &  & Macro Avg.          & 0.776 & 0.719 & \textbf{0.907} \\
        \midrule
        \multirow{4}{*}{3} & \multirow{4}{*}{Llama2 70B} 
            & Leadership         & 0.908 & \textbf{0.949} & 0.870 \\
         &  & Direct Support      & 0.352 & 0.314 & \textbf{0.400} \\
         &  & Indirect Support    & \textbf{0.514} & 0.382 & 0.789 \\
         &  & Macro Avg.          & 0.591 & 0.548 & \textbf{0.686} \\
        \midrule
        \multirow{3}{*}{4} & \multirow{3}{*}{Mistral 7x8B} 
            & Leadership         & 0.951 & \textbf{0.969} & 0.933 \\
         &  & Direct Support      & 0.580 & 0.538 & \textbf{0.629} \\
         &  & Indirect Support    & 0.783 & 0.677 & \textbf{0.926} \\
         &  & Macro Avg.          & 0.771 & 0.728 & \textbf{0.830} \\
        \bottomrule
    \end{tabular}
\end{table}

\begin{figure}[h]
    \centering
    \includegraphics[width=0.8\textwidth]{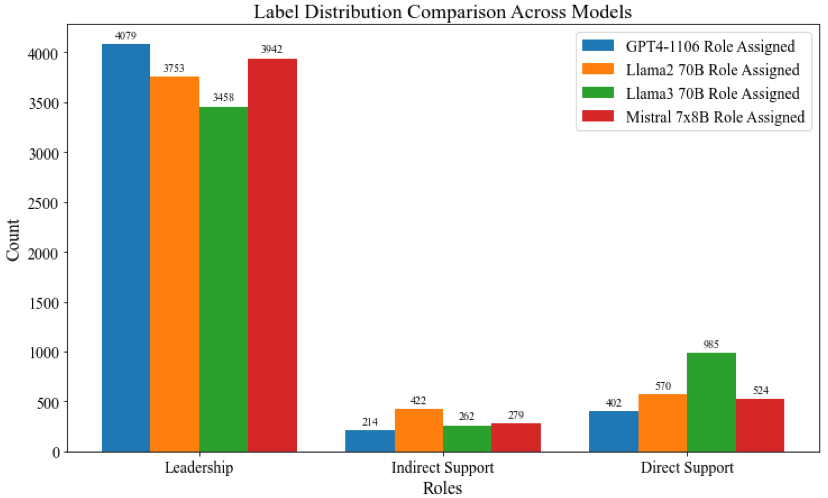} 
    \caption{Label distribution comparison across models. The figure illustrates the distribution of assigned roles—Leadership, Direct Support, and Indirect Support—by four models: GPT-4, Llama3 70B, Llama2 70B, and Mistral 7x8B.}
    \label{fig:distribution}
\end{figure}

The bar plot in Figure~\ref{fig:distribution} shows notable differences in the label distributions of the four models—GPT-4, Llama3 70B, Llama2 70B, and Mistral 7x8B—on the role classification tasks. GPT-4 predominantly assigned authors to the Leadership role more often than the other models, indicating its strong tendency to prioritize leadership activities. Conversely, Mistral 7x8B and Llama2 70B showed a higher tendency to classify authors in the Direct Support role, suggesting that these models emphasize direct support activities more than GPT-4 and Llama3 70B. For the Indirect Support role, the distribution was relatively balanced across the models, with no single model showing a significantly higher number of classifications.

Given these results, GPT-4 was selected for further role classification of the remaining data due to its higher classification accuracy. We then used GPT-4 to perform additional inference on the extended dataset to ensure comprehensive and accurate classification of author roles within scientific teams.

\subsection{Comparison of LLMs with Traditional Models}
To compare the performance of LLMs with traditional machine learning models for the task of role classification in scientific teams, we used XGBoost, a well-known and robust machine learning algorithm, as a representative of traditional models due to its strengths in classification tasks~\cite{bentejac2021comparative}.

\begin{figure}[h]
    \centering
    \includegraphics[width=0.8\textwidth]{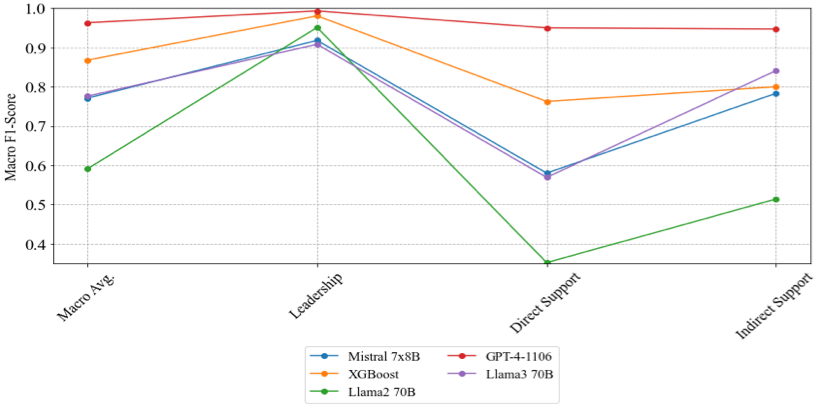} 
    \caption{Line plot for F1-score by class for each model.}
    \label{fig:results}
\end{figure}

For our comparison, XGBoost was used with a boosting learning rate of 0.2. The number of estimators was set to 1000 with a maximum tree depth of 8 for base learners. We split our dataset into training and validation sets using a stratified split based on the target variable $y$, with a ratio of 0.2 for the validation set. This stratification ensured that the distribution of classes was maintained across both sets. To transform the author contributions into features, we used the TF-IDF (Term Frequency-Inverse Document Frequency) vectorizer with a maximum of 2000 features. For label encoding, we utilized a label encoder to transform the human-corrected role labels into numerical values. The XGBoost model achieved the following scores for each role category: Direct Support: 0.76, Indirect Support: 0.8, Leadership: 0.98, and a Macro Average of 0.86, as shown in the results from Figure~\ref{fig:results}.

We also explored the performance of encoder-based transformer models, specifically BERT and RoBERTa, for this classification task. In our experiments, we fine-tuned both BERT and RoBERTa base models on our dataset using a learning rate of $2 \times 10^{-5}$. Both models were trained for 5 epochs, using a stratified split based on the target variable, with a ratio of 0.2 for the validation set. BERT achieved an average F1 score of 0.9048, and RoBERTa achieved an average F1 score of 0.8794.

\begin{table}[h]
    \centering
    \caption{Performance of BERT and RoBERTa Models on Role Classificaiton.}
    \label{tab:table2}
    \renewcommand{\multirowsetup}{\centering}
    \begin{tabular}{C{2.5cm} C{3cm} c c}
        \toprule
        \textbf{Models} & \textbf{Learning Rate} & \textbf{Epoch} & \textbf{Avg. F1-Score} \\
        \midrule
        \multirow{5}{*}{BERT-base} & \multirow{5}{*}{$2 \times 10^{-5}$}
            & 1 & 0.7927 \\
         & & 2 & 0.8560 \\
         & & \underline{3} & \underline{0.8925} \\
         & & \textbf{4} & \textbf{0.9048} \\
         & & 5 & 0.8869 \\
        \midrule
        \multirow{5}{*}{RoBERTa-base} & \multirow{5}{*}{$2 \times 10^{-5}$}
            & 1 & 0.8444 \\
         & & 2 & 0.8570 \\
         & & 3 & 0.8191 \\
         & & \underline{4} & \underline{0.8637} \\
         & & \textbf{5} & \textbf{0.8794} \\
        \bottomrule
    \end{tabular}
\end{table}

Although these models also achieved promising results, it is important to emphasize several key advantages of LLMs over traditional machine learning models. First, LLMs, such as GPT-4, can perform classification tasks without requiring any training data. This few-shot capability allows LLMs to generalize from a large body of knowledge and apply it to new, unseen tasks. Second, unlike models that rely heavily on manually labeled training datasets, LLMs can infer roles and classifications directly from text. This significantly reduces the time and effort required for data preparation. Third, LLMs have advanced natural language understanding capabilities that allow them to capture complex contexts and nuances in author contributions. This results in more accurate and contextually relevant classifications.

Despite the lack of fine-tuning, GPT-4 showed an impressive overall score, approaching an F1 score of 0.97. This result shows the potential of LLMs for role classification tasks. One key reason LLMs outperform those models is their ability to handle complex and nuanced contribution statements that may contain ambiguous language or non-standard phrasing. For example, phrases such as "contributed extensively to the work presented in this paper" or "was involved in discussions that developed the understanding of the physical processes" present challenges for rule-based approaches, which rely on specific keywords. LLMs excel at interpreting such statements by capturing the context and inferring the appropriate role classification. This deep contextual understanding allows LLMs to surpass models such as XGBoost and BERT, particularly when authors mention multiple roles without emphasizing a primary one. While machine learning models such as XGBoost and BERT have their strengths, especially in structured data classification, the inherent advantages of LLMs make them a powerful tool for understanding and categorizing scientific roles.

Therefore, our comparison shows that while models such as XGBoost and BERT perform well, LLMs offer superior performance and flexibility, especially in scenarios where labeled data is scarce or when the task requires deep contextual understanding. The use of LLMs for role classification in scientific teams thus represents a promising direction for future research and applications.

\section{Scalable Predictive Modeling for Author Role Classification}
\label{sec:role-classification-predictive-model}

To enable large-scale and efficient analysis of author roles in scientific collaborations, we address the computational challenges associated with directly applying GPT-4 for large-scale predictions. While GPT-4 demonstrates exceptional performance in classifying author roles, its inference is computationally expensive and impractical for analyzing millions of authors. To overcome this challenge, we utilize labels generated from GPT-4 to train a dense neural network, which enables efficient classification of new data with minimal computational overhead.

\subsection{Dataset Construction and Feature Engineering}
\label{subsec:dataset-feature-engineering}

To build our predictive model, we first constructed a robust dataset derived from the OpenAlex database, which provides extensive bibliometric and metadata on academic publications. We then moved on to building a predictive role classification model using feature engineering.

\begin{table}[ht]
    \centering
    \caption{Categorization of the 10 Extracted Features.}
    \label{tab:features}
    
    \renewcommand{\multirowsetup}{\centering}
    
    \begin{tabular}{
        C{1cm}   
        C{5cm}   
        C{7cm}   
    }
        \toprule
        \textbf{Index} & \textbf{Category} & \textbf{Feature Name} \\
        \midrule
        
        1 & \multirow{2}{*}{Contribution Metrics} & Contribution to References \\
        2 &                                       & Contribution to Topics      \\
        \midrule
        
        3 & \multirow{2}{*}{Leadership Metrics} & Probability of Leading \\
        4 &                                      & Probability of Leading Correspondence \\
        \midrule
        
        5 & Career Duration Metrics & Career Age \\
        \midrule
        
        6 & \multirow{2}{*}{Citation Metrics} & Citation Count \\
        7 &                                     & Citation Impact per Year \\
        \midrule
        
        8 & Research Diversity Metrics & Unique Topics \\
        \midrule
        
        9 & Publication Metrics & Total Publications \\
        \midrule
        
        10 & Collaboration Diversity Metrics & Institutional Diversity \\
        \bottomrule
    \end{tabular}
\end{table}

As shown in Table~\ref{tab:features}, we focused on 10 predictive features that capture the complex dynamics of academic authorship and contributions. Of these, eight features are derived from Xu et al~\cite{xu2022flat}. These features are:

\begin{enumerate}
    \item \textbf{Contribution to References} assesses the extent to which an author contributes references in their work.
    \[
    \mathrm{Contribution}_{(\mathrm{References})} 
    = \frac{\mathrm{Overlap}_{(\mathrm{References})}}{\mathrm{Total}_{(\mathrm{References})}}
    \]

    \item \textbf{Contribution to Topics} assesses the direction and influence of an author's previous work on the current research topic.
    \[
    \mathrm{Contribution}_{(\mathrm{Topics})}
    = \frac{\mathrm{Overlap}_{(\mathrm{Topics})}}{\mathrm{Total}_{(\mathrm{Topics})}}
    \]

    \item \textbf{Probability of Leading} indicates the likelihood that an author has often been the first author on previous papers.
    \[
    \mathrm{Probability}_{(\mathrm{Leading})}
    = \frac{\mathrm{Num}_{(\mathrm{Times\ First\ Author})}}{\mathrm{Total}_{(\mathrm{Papers})}}
    \]

    \item \textbf{Probability of Leading Correspondence} shows the probability that an author has been the corresponding author on previous papers.
    \[
    \mathrm{Probability}_{(\mathrm{Leading\ Correspondence})}
    = \frac{\mathrm{Num}_{(\mathrm{Times\ Corresponding\ Author})}}
      {\mathrm{Total}_{(\mathrm{Papers})}}
    \]

    \item \textbf{Career Age} measures the total number of years an author has been active in research.
    \[
    \mathrm{Career\ Age} 
    = \mathrm{Year}_{(\mathrm{Last\ Publication})}
      - \mathrm{Year}_{(\mathrm{First\ Publication})}
    \]

    \item \textbf{Citation Count} refers to the total number of citations an author has received for all of his or her previous publications.
    \[
    \mathrm{Citation\ Count}
    = \sum \mathrm{Citations}_{(\mathrm{All\ Publications})}
    \]

    \item \textbf{Unique Topics} is operationalized by the number of unique research topics an author has covered in his or her publication history.
    \[
    \mathrm{UniqueTopics}
    = \mathrm{Count}_{(\mathrm{Unique\ Keywords})}
    \]

    \item \textbf{Total Publications} reflects the total number of papers an author has published.
    \[
    \mathrm{Total\ Publications}
    = \sum (\mathrm{Publications})
    \]
\end{enumerate}

These eight features provide a well-rounded assessment of an author's contributions in the context of collaborative research teams. To further improve the model performance, we introduce two additional features to capture aspects of sustained research impact and collaboration diversity not fully represented by the original features:

\begin{enumerate}
    \item \textbf{Citation Impact per Year} is calculated by taking the total number of citations an author has received across all publications and dividing it by the number of years the author has been active, providing a measure of the average citation impact per year.
    \[
    \mathrm{Citation\ Impact\ per\ Year}
    = \frac{\sum \mathrm{Citation}_{(\mathrm{All\ Publications})}}
      {\mathrm{Years}_{(\mathrm{Active})}}
    \]

    \item \textbf{Institutional Diversity} measures the number of unique institutions an author has been affiliated with or collaborated with over their career. A higher diversity of institutions suggests a broader network and potentially more diverse research experiences and influences.
    \[
    \mathrm{Institutional\ Diversity}
    = \mathrm{Count}_{(\mathrm{Unique\ Institutions})}
    \]
\end{enumerate}

These 10 features were selected based on their relevance and ability to comprehensively capture the various dimensions of an author's contributions in collaborative research settings. \textit{Contribution to References} and \textit{Topics} highlight intellectual contributions by assessing how previous work influences current research. \textit{Probability of Leading} and \textit{Probability of Leading Correspondence} indicate typical roles and leadership responsibilities in research projects. \textit{Career Age} measures the length of an author's research career and provides context for their experience. \textit{Citation Count} directly measures the impact and recognition of an author's work. \textit{Unique Topics} and \textit{Total Publications} assess the breadth and productivity of an author's research interests.

We introduced two additional features: \textit{Citation Impact per Year} and \textit{Institutional Diversity}, to capture aspects of sustained research impact and collaboration diversity not fully represented by the original features. \textit{Citation Impact per Year} normalizes citation counts by years of activity, indicating sustained impact. \textit{Institutional Diversity} captures the extent of collaboration across different institutions, reflecting a broad research network and diverse research exposure. The combination of these features enables the model to effectively classify authors' roles, ensuring a balanced and comprehensive assessment of their contributions.

\subsection{Data Splitting and Normalization}
\label{subsec:data-splitting}

To ensure a robust evaluation of the prediction models, we expanded our dataset by further inferring the roles of 10{,}000 distinct authors using the classification framework developed in the first phase, resulting in a total dataset of about 15{,}000 data points. Originally, the dataset included three classes: Leadership, Direct Support, and Indirect Support. For the purpose of this \emph{binary classification} task, we merged the Direct Support and Indirect Support roles into a single ``Support'' category, while maintaining ``Leadership'' as a distinct category. This expanded dataset included detailed role classifications based on each author's contributions. The dataset was then divided into training and test sets using stratified sampling to maintain the distribution of the target variable. A ratio of 0.8 to 0.2 was used for the split, stratifying by two types of roles classified as leadership and non-leadership types.

Additionally, Normalizing the dataset was a critical step in preparing it for predictive modeling. We applied a min-max scaling to each feature to ensure that they contributed equally to the model, without any single feature disproportionately influencing the results due to its size. The formula used for min-max normalization is given by:
\[
\mathrm{Normalized\ Value}
= \frac{\mathrm{Value} - \mathrm{Min}}{\mathrm{Max} - \mathrm{Min}}
\]

\subsection{Predictive Model Training and Performance}
\label{subsec:model-training-results}

For the role classification task, we constructed a Dense Neural Network (DNN) model. The architecture of the DNN was designed to effectively process the 10 predictive features and accurately classify author roles. We chose this model due to its simplicity and adequacy for our task. Dense neural networks are straightforward to implement and require less computational power compared to more complex architectures, making them suitable for large-scale applications. The model can process large datasets quickly, which is essential for applying the classification to extensive corpora of scientific publications.

\[
\hat{y} = f(x)
= \sigma_3 \Bigl( W_3 \cdot \sigma_2 \bigl( W_2 \cdot \sigma_1 ( W_1 \cdot x + b_1 ) + b_2 \bigr) + b_3 \Bigr)
\]
where:
\begin{itemize}
    \item $x$ represents the input vector of the predictive features.
    \item $W_1, W_2, W_3$ are the weight matrices for the hidden layers and output layer.
    \item $\sigma_1, \sigma_2, \sigma_3$ are the ReLU (Rectified Linear Unit) activation functions for the hidden layers and a Sigmoid activation function for the output layer.
    \item $b_1, b_2, b_3$ are the bias vectors for the hidden layers and the output layer.
\end{itemize}

Initially, the model was trained using only the first eight features. This version of the model achieved an F1 score of about 0.74. To improve model performance, we introduced two additional features: (1) \textit{Citation Impact per Year} and (2) \textit{Institutional Diversity}. With the inclusion of these features, the model's F1 score improved to approximately 0.76. This demonstrates the positive impact of these additional features on the accuracy of our role classification model. The model was trained for 20 epochs, and its performance was optimized through hyperparameter tuning. By ensuring a balanced and representative sample, we have laid a solid foundation for accurate and reliable author role classification.

We also experimented with the same dataset using XGBoost, configured with a boosting learning rate of 0.2. The number of estimators was set to 1000, and the maximum tree depth was set to 8 for base learners. The XGBoost model achieved a slightly better F1 score of 0.78 compared to the dense neural network. However, considering our goal to develop a model suitable for larger datasets, we chose the dense neural network due to its simplicity and scalability.

Furthermore, in the study by Xu et al., they achieved an F1 score of approximately 0.79 by utilizing a larger dataset and more complex modeling techniques. In contrast, our study achieved an F1 score of 0.76 using a smaller dataset and a relatively simple dense neural network model. This demonstrates that our approach is both robust and efficient, achieving comparable performance with less computational complexity and resource investment.

\subsection{Interpretability Analysis Using SHAP}

To further validate feature importance in our classification task, we apply SHAP (SHapley Additive exPlanations)~\cite{lundberg2017unified,lundberg2018consistent} to investigate how each predictive feature contributes to the model's decision-making process. Specifically, we utilize \textit{Gradient SHAP}, which efficiently estimates feature attributions for deep learning models using gradient-based sampling. the SHAP value for a feature $i$ is calculated as:

\[
\phi_i = \sum_{S \subseteq F \\ i \notin S} \frac{|S|! (M - |S| - 1)!}{M!} \left[ f(S \cup \{i\}) - f(S) \right]
\]

where:
\begin{itemize}
    \item $F$ is the set of all features,
    \item $S$ is a subset of features excluding $i$,
    \item $f(S)$ is the model's prediction using only the features in $S$,
    \item $\phi_i$ is the SHAP value for feature $i$.
\end{itemize}

\begin{figure}[h]
    \centering
    \includegraphics[width=0.8\textwidth]{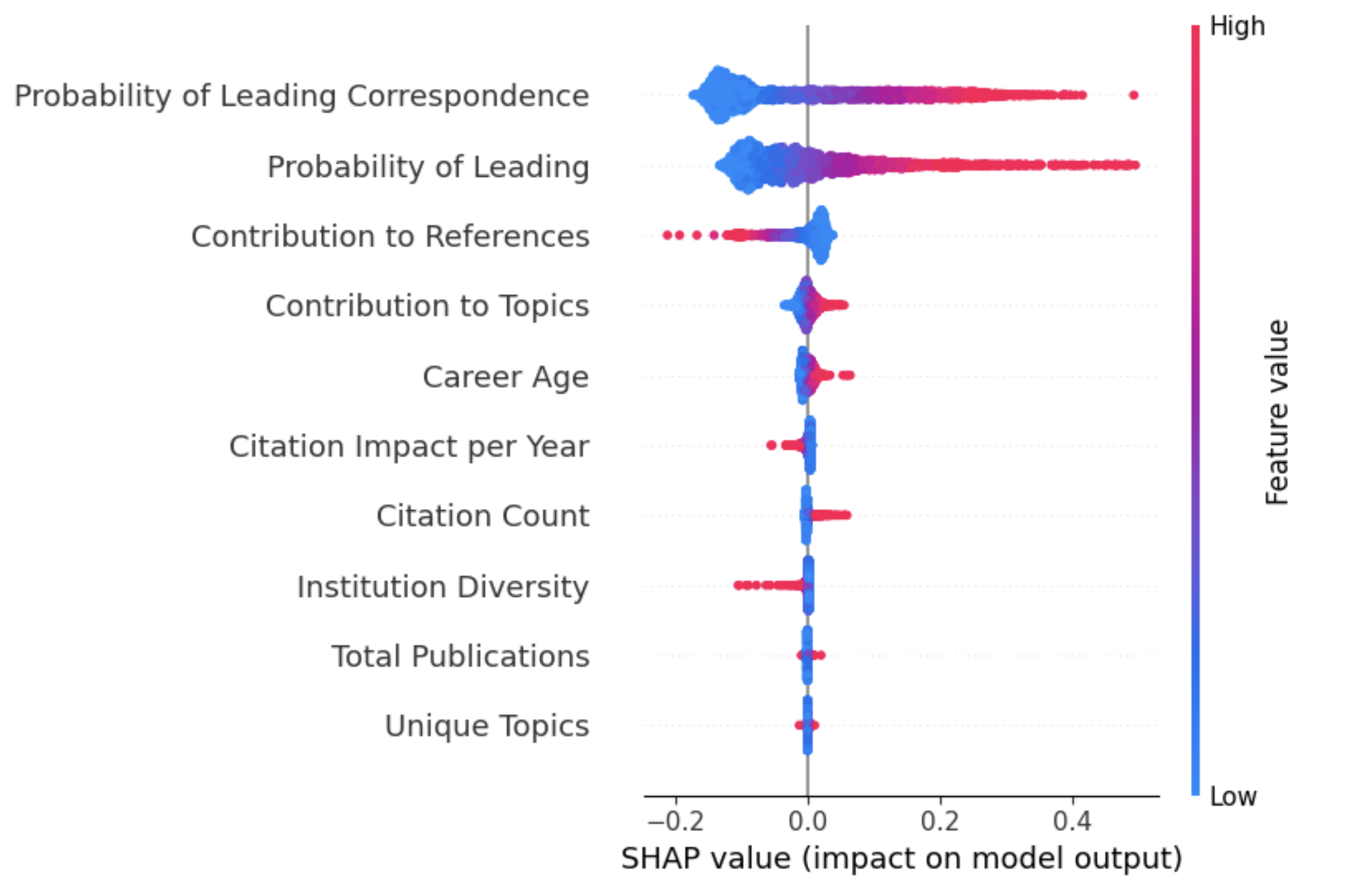}
    \caption{SHAP summary plot showing feature importance and the directional impact of each feature on the model's predictions.}
    \label{fig:shap}
\end{figure}

As shown in Figure~\ref{fig:shap}, \textit{Probability of Leading Correspondence} and \textit{Probability of Leading} emerge as the most influential features in distinguishing leadership roles. These features likely capture the responsibility and visibility of an author within a research project, suggesting that consistent first or corresponding authorship is a strong indicator of leadership roles. Their high SHAP values further confirm their direct impact on classification predictions, reinforcing their centrality in the model’s decision-making process.

Subsequently, \textit{Contribution to References} and \textit{Contribution to Topics} also hold substantial importance, showcasing their role in capturing intellectual contributions. A higher contribution to references indicates an author's influence on shaping discussions within a paper, while topic contributions suggest an author's role in guiding the thematic direction of research. The positive SHAP impact of these features suggests that intellectual involvement plays a critical role in distinguishing different author roles.

Additionally, the newly proposed features in this paper, including: \textit{Citation Impact per Year} and \textit{Institutional Diversity}, provide meaningful contributions. \textit{Citation Impact per Year} accounts for the sustained influence of an author’s research over time, making it a valuable indicator of long-term academic impact. Its moderate importance in the SHAP analysis suggests that while citations are significant, they are not the sole determinant of leadership roles. \textit{Institutional Diversity}, on the other hand, captures the extent of an author's research collaborations across different institutions. A higher value for this feature suggests exposure to diverse academic environments, which may contribute to broader collaborative roles rather than direct leadership.

\section{Conclusions}
\label{sec:conclusions}

In this paper, we have presented a transformative approach to classifying author roles within scientific teams using advanced LLMs such as GPT-4, Llama3 70B, Llama2 70B, and Mistral 7x8B. By integrating few-shot prompting and extensive feature engineering, we categorized author roles into Leadership, Direct Support, and Indirect Support. Our results showed that GPT-4 achieved the highest accuracy across multiple categories, outperforming other models in role classification tasks. By employing these advanced LLMs, we have improved the understanding and analysis of team roles in scientific research, providing a more detailed and accurate classification than traditional models.

We developed a predictive deep learning model incorporating ten features that capture the complex dynamics of academic authorship and contributions. These features included contribution to references, contribution to topics, probability of leading, probability of managing correspondence, career age, number of citations, unique topics, total publications, citation impact per year, and institutional diversity. The model, trained on a dataset derived from the OpenAlex database and self-reported contribution data, achieved an F1 score of 0.76, indicating robust performance in classifying author roles with a Dense Neural Network (DNN).

In addition, this methodology allows us to further calculate the L-Ratio, which quantifies the proportion of leadership roles within research teams. In future work, by integrating the L-Ratio with other variables, we can explore the relationship between leadership dynamics and various aspects of academic performance. For instance, using the L-Ratio to analyze the long-term performance of junior team members can provide insights into their career trajectories and contributions over time. This analysis aims to highlight the importance of leadership dynamics in scientific collaboration, providing valuable information on how junior researchers’ roles and impacts evolve within academic teams.

Overall, our approach not only improves the accuracy of author role classification, but also provides deeper insights into the complex dynamics of scientific teamwork. This work represents a significant step forward in understanding contributions and collaborations within scientific teams, thereby improving our ability to analyze and promote effective scientific research environments.

\section{Discussion: Technical and Practical Implications}
Our study introduces a technical framework that leverages few-shot prompting with LLMs and a scalable neural network to generate high-quality labels without extensive training data. The use of SHAP for interpretability highlights the key factors—such as the probability of leading—that define leadership roles. Practically, our approach supports a deeper understanding of scientific collaboration. Moreover, the scalable predictive model paves the way for automated monitoring of author contributions across large datasets.

Despite these promising results, our study has limitations that suggest directions for future research. While advanced LLMs such as GPT-4 demonstrate superior performance in author role classification, their proprietary nature and computational requirements may limit accessibility for some researchers. Additionally, our reliance on data derived from the OpenAlex database may not capture the full diversity of author contributions across all scientific discipline and publication venues. To address these limitations, future work could involve exploring zero-shot inference with LLMs to obtain more natural and pattern-rich results. Furthermore, analyzing the patterns and insights derived from zero-shot inference could enhance our understanding of author roles and contributions, particularly in handling complex or ambiguous cases that challenge traditional rule-based approaches.

\section{Data Availability}
\label{sec:data-availability}

The self-reported data analyzed in this paper is available at 
\url{https://github.com/fenglixu/Self-report-Contribution-Data}. All data used in this study are open and available at OpenAlex 
\url{https://openalex.org/}.

\section{Author Contribution}

\textbf{W.S.} contributed to Conceptualization, Data curation, Formal analysis, Investigation, Methodology, Software, and Writing – original draft.

\textbf{Y.B.} contributed to Conceptualization, Methodology, Project administration, Resources, Supervision, and Writing – review \& editing.

Both authors \textbf{W.S.} and \textbf{Y.B.} contributed equally to the study design, methodology development, and manuscript preparation. All authors read and approved the final manuscript.

\section{Competing interests}
The authors do not declare any competing interests.

\section{Acknowledgments}
\label{sec:acknowledgments}

Yi Bu acknowledges the support from the National Natural Science Foundation of China (\#72474009, \#72104007, and \#72174016) and from the 2024 Cultural Research Project of Ningbo under Grant WH24-2-4. The authors are grateful to Yifan Tian and Zonghao Yuan who had fruitful discussions with the authors.

\bibliographystyle{unsrt}  
\bibliography{references}  

\begin{thebibliography}{10}

\bibitem{wuchty2007increasing}
Stefan Wuchty, Benjamin~F Jones, and Brian Uzzi.
\newblock The increasing dominance of teams in production of knowledge.
\newblock {\em Science}, 316(5827):1036--1039, 2007.

\bibitem{xu2022flat}
Fengli Xu, Lingfei Wu, and James Evans.
\newblock Flat teams drive scientific innovation.
\newblock {\em Proceedings of the National Academy of Sciences}, 119(23):e2200927119, 2022.

\bibitem{xu2024ai}
Ruoxi Xu, Yingfei Sun, Mengjie Ren, Shiguang Guo, Ruotong Pan, Hongyu Lin, Le~Sun, and Xianpei Han.
\newblock Ai for social science and social science of ai: A survey.
\newblock {\em Information Processing \& Management}, 61(3):103665, 2024.

\bibitem{krenn2020predicting}
Mario Krenn and Anton Zeilinger.
\newblock Predicting research trends with semantic and neural networks with an application in quantum physics.
\newblock {\em Proceedings of the National Academy of Sciences}, 117(4):1910--1916, 2020.

\bibitem{achiam2023gpt}
Josh Achiam, Steven Adler, Sandhini Agarwal, Lama Ahmad, Ilge Akkaya, Florencia~Leoni Aleman, Diogo Almeida, Janko Altenschmidt, Sam Altman, Shyamal Anadkat, et~al.
\newblock Gpt-4 technical report.
\newblock {\em arXiv preprint arXiv:2303.08774}, 2023.

\bibitem{jiang2024mixtral}
Albert~Q Jiang, Alexandre Sablayrolles, Antoine Roux, Arthur Mensch, Blanche Savary, Chris Bamford, Devendra~Singh Chaplot, Diego de~las Casas, Emma~Bou Hanna, Florian Bressand, et~al.
\newblock Mixtral of experts.
\newblock {\em arXiv preprint arXiv:2401.04088}, 2024.

\bibitem{touvron2023llama}
Hugo Touvron, Thibaut Lavril, Gautier Izacard, Xavier Martinet, Marie-Anne Lachaux, Timoth{\'e}e Lacroix, Baptiste Rozi{\`e}re, Naman Goyal, Eric Hambro, Faisal Azhar, et~al.
\newblock Llama: Open and efficient foundation language models.
\newblock {\em arXiv preprint arXiv:2302.13971}, 2023.

\bibitem{priem2022openalex}
Jason Priem, Heather Piwowar, and Richard Orr.
\newblock Openalex: A fully-open index of scholarly works, authors, venues, institutions, and concepts.
\newblock {\em arXiv preprint arXiv:2205.01833}, 2022.

\bibitem{lundberg2017unified}
Scott Lundberg.
\newblock A unified approach to interpreting model predictions.
\newblock {\em arXiv preprint arXiv:1705.07874}, 2017.

\bibitem{uzzi2013atypical}
Brian Uzzi, Satyam Mukherjee, Michael Stringer, and Ben Jones.
\newblock Atypical combinations and scientific impact.
\newblock {\em Science}, 342(6157):468--472, 2013.

\bibitem{katz1997research}
J~Sylvan Katz and Ben~R Martin.
\newblock What is research collaboration?
\newblock {\em Research policy}, 26(1):1--18, 1997.

\bibitem{melin1996studying}
G{\"o}ran Melin and Olle Persson.
\newblock Studying research collaboration using co-authorships.
\newblock {\em Scientometrics}, 36:363--377, 1996.

\bibitem{hollenbeck2012beyond}
John~R Hollenbeck, Bianca Beersma, and Maartje~E Schouten.
\newblock Beyond team types and taxonomies: A dimensional scaling conceptualization for team description.
\newblock {\em Academy of Management Review}, 37(1):82--106, 2012.

\bibitem{woolley2010evidence}
Anita~Williams Woolley, Christopher~F Chabris, Alex Pentland, Nada Hashmi, and Thomas~W Malone.
\newblock Evidence for a collective intelligence factor in the performance of human groups.
\newblock {\em science}, 330(6004):686--688, 2010.

\bibitem{wu2019large}
Lingfei Wu, Dashun Wang, and James~A Evans.
\newblock Large teams develop and small teams disrupt science and technology.
\newblock {\em Nature}, 566(7744):378--382, 2019.

\bibitem{cummings2005collaborative}
Jonathon~N Cummings and Sara Kiesler.
\newblock Collaborative research across disciplinary and organizational boundaries.
\newblock {\em Social studies of science}, 35(5):703--722, 2005.

\bibitem{anicich2015hierarchical}
Eric~M Anicich, Roderick~I Swaab, and Adam~D Galinsky.
\newblock Hierarchical cultural values predict success and mortality in high-stakes teams.
\newblock {\em Proceedings of the National Academy of Sciences}, 112(5):1338--1343, 2015.

\bibitem{haeussler2020division}
Carolin Haeussler and Henry Sauermann.
\newblock Division of labor in collaborative knowledge production: The role of team size and interdisciplinarity.
\newblock {\em Research Policy}, 49(6):103987, 2020.

\bibitem{waltman2012new}
Ludo Waltman and Nees~Jan Van~Eck.
\newblock A new methodology for constructing a publication-level classification system of science.
\newblock {\em Journal of the American Society for Information Science and Technology}, 63(12):2378--2392, 2012.

\bibitem{boyack2010co}
Kevin~W Boyack and Richard Klavans.
\newblock Co-citation analysis, bibliographic coupling, and direct citation: Which citation approach represents the research front most accurately?
\newblock {\em Journal of the American Society for information Science and Technology}, 61(12):2389--2404, 2010.

\bibitem{blondel2008fast}
Vincent~D Blondel, Jean-Loup Guillaume, Renaud Lambiotte, and Etienne Lefebvre.
\newblock Fast unfolding of communities in large networks.
\newblock {\em Journal of statistical mechanics: theory and experiment}, 2008(10):P10008, 2008.

\bibitem{glanzel2004analysing}
Wolfgang Gl{\"a}nzel and Andr{\'a}s Schubert.
\newblock Analysing scientific networks through co-authorship.
\newblock In {\em Handbook of quantitative science and technology research: The use of publication and patent statistics in studies of S\&T systems}, pages 257--276. Springer, 2004.

\bibitem{cheng2024method}
Zhe Cheng, Yihuan Zou, and Yueyang Zheng.
\newblock A method for identifying different types of university research teams.
\newblock {\em Humanities and Social Sciences Communications}, 11(1):1--15, 2024.

\bibitem{fujimoto2016team}
Manabu Fujimoto.
\newblock Team roles and hierarchic system in group discussion.
\newblock {\em Group Decision and Negotiation}, 25:585--608, 2016.

\bibitem{conger1998qualitative}
Jay~A Conger.
\newblock Qualitative research as the cornerstone methodology for understanding leadership.
\newblock {\em The Leadership Quarterly}, 9(1):107--121, 1998.

\bibitem{holbrook2017peer}
J~Britt Holbrook.
\newblock Peer review, interdisciplinarity, and serendipity.
\newblock 2017.

\bibitem{endersby1996collaborative}
James~W Endersby.
\newblock Collaborative research in the social sciences: Multiple authorship and publication credit.
\newblock {\em Social Science Quarterly}, pages 375--392, 1996.

\bibitem{brand2015beyond}
Amy Brand, Liz Allen, Micah Altman, Marjorie Hlava, and Jo~Scott.
\newblock Beyond authorship: Attribution, contribution, collaboration, and credit.
\newblock {\em Learned Publishing}, 28(2), 2015.

\bibitem{brown2020language}
Tom Brown, Benjamin Mann, Nick Ryder, Melanie Subbiah, Jared~D Kaplan, Prafulla Dhariwal, Arvind Neelakantan, Pranav Shyam, Girish Sastry, Amanda Askell, et~al.
\newblock Language models are few-shot learners.
\newblock {\em Advances in neural information processing systems}, 33:1877--1901, 2020.

\bibitem{vaswani2017attention}
A~Vaswani.
\newblock Attention is all you need.
\newblock {\em Advances in Neural Information Processing Systems}, 2017.

\bibitem{sherstinsky2020fundamentals}
Alex Sherstinsky.
\newblock Fundamentals of recurrent neural network (rnn) and long short-term memory (lstm) network.
\newblock {\em Physica D: Nonlinear Phenomena}, 404:132306, 2020.

\bibitem{liu2019roberta}
Yinhan Liu.
\newblock Roberta: A robustly optimized bert pretraining approach.
\newblock {\em arXiv preprint arXiv:1907.11692}, 364, 2019.

\bibitem{kenton2019bert}
Jacob Devlin Ming-Wei~Chang Kenton and Lee~Kristina Toutanova.
\newblock Bert: Pre-training of deep bidirectional transformers for language understanding.
\newblock In {\em Proceedings of naacL-HLT}, volume~1, page~2. Minneapolis, Minnesota, 2019.

\bibitem{raffel2020exploring}
Colin Raffel, Noam Shazeer, Adam Roberts, Katherine Lee, Sharan Narang, Michael Matena, Yanqi Zhou, Wei Li, and Peter~J Liu.
\newblock Exploring the limits of transfer learning with a unified text-to-text transformer.
\newblock {\em Journal of machine learning research}, 21(140):1--67, 2020.

\bibitem{radford2018improving}
Alec Radford.
\newblock Improving language understanding by generative pre-training.
\newblock 2018.

\bibitem{thelwall2024evaluating}
M~Thelwall.
\newblock Evaluating research quality with large language models: an analysis of chatgpt’s effectiveness with different settings and inputs.
\newblock {\em Journal of Data and Information Science}, 2024.

\bibitem{thelwall2024fields}
Mike Thelwall and Abdallah Yaghi.
\newblock In which fields can chatgpt detect journal article quality? an evaluation of ref2021 results.
\newblock {\em arXiv preprint arXiv:2409.16695}, 2024.

\bibitem{wu2024scientific}
Mengjia Wu, Gunnar Sivertsen, Lin Zhang, Fan Qi, and Yi~Zhang.
\newblock Scientific progress or societal progress? a language modelbased classification of the aims of the research in scientific publications.
\newblock {\em A Language Modelbased Classification of the Aims of the Research in Scientific Publications (April 22, 2024)}, 2024.

\bibitem{aydin2022openai}
{\"O}mer Ayd{\i}n and Enis Karaarslan.
\newblock Openai chatgpt generated literature review: Digital twin in healthcare.
\newblock {\em Ayd{\i}n, {\"O}., Karaarslan, E.(2022). OpenAI ChatGPT Generated Literature Review: Digital Twin in Healthcare. In {\"O}. Ayd{\i}n (Ed.), Emerging Computer Technologies}, 2, 2022.

\bibitem{gilardi2023chatgpt}
Fabrizio Gilardi, Meysam Alizadeh, and Ma{\"e}l Kubli.
\newblock Chatgpt outperforms crowd workers for text-annotation tasks.
\newblock {\em Proceedings of the National Academy of Sciences}, 120(30):e2305016120, 2023.

\bibitem{wang2023can}
Shuai Wang, Harrisen Scells, Bevan Koopman, and Guido Zuccon.
\newblock Can chatgpt write a good boolean query for systematic review literature search?
\newblock In {\em Proceedings of the 46th International ACM SIGIR Conference on Research and Development in Information Retrieval}, pages 1426--1436, 2023.

\bibitem{rivas2023marketing}
Pablo Rivas and Liang Zhao.
\newblock Marketing with chatgpt: Navigating the ethical terrain of gpt-based chatbot technology.
\newblock {\em AI}, 4(2):375--384, 2023.

\bibitem{bentejac2021comparative}
Candice Bent{\'e}jac, Anna Cs{\"o}rg{\H{o}}, and Gonzalo Mart{\'\i}nez-Mu{\~n}oz.
\newblock A comparative analysis of gradient boosting algorithms.
\newblock {\em Artificial Intelligence Review}, 54:1937--1967, 2021.

\bibitem{lundberg2018consistent}
Scott~M Lundberg, Gabriel~G Erion, and Su-In Lee.
\newblock Consistent individualized feature attribution for tree ensembles.
\newblock {\em arXiv preprint arXiv:1802.03888}, 2018.

\end{thebibliography}







\end{document}